\documentclass[%
 reprint,
 amsmath,amssymb,
 aps,
groupedaddress,
superscriptaddress,
]{revtex4-2}

\usepackage{graphicx}
\usepackage{subfig}
\usepackage{mathrsfs}
\usepackage{amsmath}
\usepackage{placeins}
\usepackage{epstopdf, epsfig}
\usepackage[export]{adjustbox}
\usepackage[%
    colorlinks=true,
    pdfborder={0 0 0},
    linkcolor=blue,
 citecolor=blue,
urlcolor=blue
]{hyperref}
\usepackage[abs]{overpic}
\usepackage{tikz}
\usepackage{dcolumn}% Align table columns on decimal point
\usepackage{bm}% bold math
\usepackage{multirow}
\usepackage{makecell}

\begin{document}

%\preprint{}

\title[]{Hydrodynamic interactions and extreme particle clustering in turbulence}

\date{\today}

\author{Andrew D. Bragg}
 \email{andrew.bragg@duke.edu}
 \affiliation{Department of Civil and Environmental Engineering, Duke University, Durham, NC 27708, USA}%

\author{Adam L. Hammond}
\thanks{The first two authors have equal contribution.}
\affiliation{Department of Mechanical and Aerospace Engineering, University at Buffalo, Buffalo, NY,
14260, United States
}%

\author{Rohit Dhariwal}
\affiliation{%
 Center for Institutional Research Computing Staff, Washington State University, Pullman, WA
}%
\author{Hui Meng}
 \email{huimeng@buffalo.edu}
\affiliation{Department of Mechanical and Aerospace Engineering, University at Buffalo, Buffalo, NY,
14260, United States
}%

\begin{abstract}
From new detailed experimental data, we found that the Radial Distribution Function (RDF) of inertial particles in turbulence grows explosively with $r^{-6}$ scaling as the collision radius is approached. We corrected a theory by Yavuz et al. (Phys. Rev. Lett. 120, 244504 (2018)) based on hydrodynamic interactions between pairs of weakly inertial particles, and demonstrate that even this corrected theory cannot explain the observed RDF behavior. We explore several alternative mechanisms for the discrepancy that were not included in the theory and show that none of them are likely the explanation, suggesting new, yet to be identified physical mechanisms are at play.
\end{abstract}%

\maketitle

Small, inertial particles can spontaneously cluster in incompressible turbulent flows, an effect considered important for droplet collision rates in atmospheric clouds \cite{shaw03,grabowski13} and planetesimal formation in  turbulent circumstellar disks \cite{johansen07}. However, even in the absence of particle inertia, hydrodynamic interactions (HI) between pairs of particles can also lead to particle clustering \cite{bkl97}. This behavior has been explored theoretically for inertia-free particles ($St=0$, where Stokes number $St$ defined as the particle response time $\tau_p$ divided by the Kolmogorov timescale $\tau_\eta$) in low Reynolds number flows \cite{batchelor72} as well as turbulent flows \cite{bkl97}. These analyses show that the Radial Distribution Function (RDF) $g(r)$, which quantifies clustering, scales according to separation $r$ as $g(r) \sim r^{-6}$, for $r$ exceeding a few multiples of the particle diameter (i.e. the ``far-field'' regime). 

Because HI occur over scales on the order of the particle size \cite{bkl97}, which is often on the order of microns, it is challenging to experimentally observe the effects of HI on particle clustering in turbulent flows. Indeed, until recently such experimental inquiry was prohibited due to spatiotemporal resolution and perspective overlap preventing the identification of particle pairs with very small separations \cite{hammond21,kearney2020lagrangian}. Yavuz \emph{et al} \cite{yavuz18} presented the first experimental evidence of extreme inertial particle clustering as $r$ approaches the collision radius. By extending the analysis of the inertia-free theory by Brunk \emph{et al} \cite{bkl97} to the case of weakly inertial particles ($St\ll 1$) using the drift-diffusion model of Chun \emph{et al} \cite{chun05}, Yavuz \emph{et al} claimed that this RDF enhancement was due to the combined effect of particle inertia and particle-pair HI \cite{yavuz18}. Their resulting theory predicted that $g(r)$ would scale as $r^{-6}$ for a regime of $r$; however, they did not observe this scaling in their experiments. They speculated that the $r^{-6}$ regime must occur at a scale smaller than they were able to reliably measure. Their theory did, however, predict a contribution to the RDF arising from particle inertia that appears at slightly larger separations, and they claimed that this contribution explained the extreme clustering observed in their experiments. However, as we demonstrate, the theory by Yavuz \emph{et al} \cite{yavuz18} contains a number of errors. When these errors are corrected, the theory actually indicates that the inertial contribution to HI hinders clustering instead of enhancing it. Therefore, this theory by Yavuz \emph{et al} \cite{yavuz18} cannot explain the extreme clustering they observed. In addition, their measurement data exhibited significant scatter at the small-$r$ region where the RDF increased explosively, which does not inspire confidence in the new data they offer.

Recently, Hammond and Meng \cite{hammond21} reported a new experimental RDF measurement of inertial particles ($St=0.74$, and $a=14.25\mu m$, where $a$ is particle radius) in isotropic turbulence at $r$ down to near-contact ($r/a = 2.07$). Using a novel particle tracking approach based on four-pulse Shake-the-Box (4P-STB), they obtained RDF and relative velocity results at unprecedentedly high resolution. When $r$ went below $r/\eta = O(1)$ ($\eta$: Kolmogorov length) corresponding to $r/a= O(10)$, their measured $g(r)$ grew explosively due to particle-particle interactions. The order of magnitude of $g(r)$ at near-contact matched that of Yavuz \emph{et al} \cite{yavuz18}; however, their improved measurement resolution clearly showed that $g(r)$ scaled as $r^{-6}$ in this regime. This scaling is reminiscent of the prediction for the RDF of inertia-free particles subject to HI \cite{bkl97}, hinting that HI may have driven the observed explosive clustering. In order to understand the mechanics behind these observed extreme clustering data, more theoretical analysis is required. Moreover, to validate the theory, more experimental data is desired.

In this letter, we present additional experimental RDF measurements over a range of $St$ (0.07 to 1.06) and particle radius $a$ ($3.75$ to $20.75\mu m$), closely examine the theoretical analysis by Yavuz \emph{et al} \cite{yavuz18}, correct errors contained therein, and compare the scaling exponents predicted by the corrected theory with those of the new experimental dataset. We found enormous discrepancies between the theoretical predictions and experimental results. Therefore, we also investigated potential explanations for these discrepancies.

\textbf{Theory -} We now derive a solution for $g(r)$ while providing a physical explanation for how HI leads to particle clustering in turbulence. As shown in the Supplemental Material, the steady-state RDF $g(r)$ may be expressed exactly as (with $t\to \infty$) \cite{tom19}
\begin{align}
g(r)=\Big\langle\exp\Big[-\int^t_0\bm{\nabla\cdot \mathcal{W}}(\bm{\xi}(s),s)\,ds\Big]\Big\rangle_{{r}},\label{rhosol}
\end{align}
where $\bm{\mathcal{W}}$ is the relative velocity between two particles, $\partial_s\bm{\xi\equiv  \mathcal{W}}$, and $\langle\cdot\rangle_{{r}}$ denotes an ensemble average conditioned on the particles having the separation $\|\bm{\xi}(t)\|={r}$. For fluid particles, $\bm{\nabla\cdot \mathcal{W}}=0$ in an incompressible flow and so $g(r)=1$, i.e. they do not cluster. However, if $\bm{\nabla\cdot \mathcal{W}}$ is finite, clustering may occur with $g(r)>1$.

If we consider monodisperse particle-pairs with radius $a$ that experience HI, then for  $St\to0$ we have $\bm{\nabla\cdot }\bm{\mathcal{W}}=\lambda\mathcal{S}_{\parallel}$ \cite{bkl97}, where $\lambda\geq 0$ is a non-dimensional, nonlinear function of $r/a$ that characterizes the HI, and $\mathcal{S}_{\parallel}$ is the fluid strain-rate parallel to the particle-pair separation vector. Since $\lambda\geq 0$, then the particle field will be compressed in regions where $\mathcal{S}_{\parallel}<0$, and dilated in regions where $\mathcal{S}_{\parallel}>0$. That $\bm{\nabla\cdot }\bm{\mathcal{W}}\neq 0$ is due to the disturbance fields in the flow produced by displacement of the fluid around the two particles, which in turn generates forces on the particles. This force either causes the particles to be attracted or repelled from each other, and vanishes for fluid particles ($a=0$) since they do not disturb the flow.

Using $\bm{\nabla\cdot }\bm{\mathcal{W}}=\lambda\mathcal{S}_{\parallel}$ in \eqref{rhosol} we see that $g(r)>1$ is associated with a preference for trajectories with $\int^t_0 \lambda(\xi(s))\mathcal{S}_{\parallel}(s)\,ds<0$, that arises precisely because the particles are compressed into regions where $\mathcal{S}_{\parallel}<0$. This phenomenon is similar to the case of inertial particles with $St\ll1$ (without HI) whose clustering is driven by preferential sampling of weak-vorticity, high-strain regions of the flow \cite{chun05,bragg14b,bragg14d}, that arises due to the particles being centrifuged out of vortical regions of the flow \cite{maxey87}.

The HI effect on clustering is dependent on $St$. Since HI only occur when $r$ is sufficiently small, we define $\ell_a$ as the lengthscale of the hydrodynamic disturbance, below which HI become appreciable. At $r>\ell_a$, HI are not important, and the clustering arises solely due to how inertia modifies the particle interaction with the turbulence \cite{bragg14b,bragg14d}. For $r<\ell_a$ and $St\ll1$, the physical mechanism leading to RDF enhancement comes from particles being compressed into regions where $\mathcal{S}_{\parallel}<0$ as discussed above, with sub-leading corrections to the trajectories due to inertia. For $r<\ell_a$ and $St\geq \mathcal{O}(1)$, the mechanism generating $g(r\leq\ell_a)>1$ will be strongly affected by the non-local dependence of $\bm{\mathcal{W}}(\bm{\xi}(s),s)$ upon the turbulence the particles have experienced along their path-history at times $s'<s$ \cite{bragg14b,bragg14d}.

While \eqref{rhosol} is useful for understanding how particles cluster, it is not straightforward to derive from this a closed expression for $g(r)$. Yavuz \emph{et al} \cite{yavuz18} developed a theoretical model for $g(r)$ in the regime $St\ll1$, based on the drift-diffusion models of \cite{bkl97,chun05}. The model assumes monodisperse particle-pairs that experience HI and Stokes drag forces, suspended in a turbulent flow. Unfortunately, we have found several errors in their analysis. In the Supplemental Material we derive in detail the correct version of the theory, confining attention to the far-field asymptotic behavior ($r\gg a$, although in practice the far-field asymptotics are valid down to $r/a\approx 2.05$ \cite{bkl97}), and retaining terms up to order $St^2$, the same regime considered in \cite{yavuz18}. This gives rise to
\begin{align}
g(r)\sim\Big(\frac{r}{a}\Big)^{-St^2\mu_4} \exp\Big(\mu_1 \frac{a^6}{r^6} +(St\mu_2 +St^2\mu_3)\frac{a}{r}\Big),
\label{RDF_predic}
\end{align}
where explicit forms of the coefficients $\mu_1,\mu_2,\mu_3,\mu_4$ are given in the Supplemental Material. 

Two of the four terms in \eqref{RDF_predic} have been characterized in the literature previously. In the absence of HI, ${\mu_1=\mu_2=\mu_3=0}$, and $g(r)\sim (r/a)^{- St^2\mu_4}$ describes the clustering due solely to particle inertia \cite{chun05}. The leading HI contribution $\exp( \mu_1 a^6/r^6)$ is the far-field form of the result derived in Brunk \emph{et al} \cite{bkl97}, which is independent of $St$ and describes the clustering due to HI that can occur even for $St=0$. 

The $\mathcal{O} (St^2)$ inertial contribution to the clustering arising from HI, $\exp(St^2 \mu_3 a/r)$ was first derived in \cite{yavuz18}, where they determined $\mu_3$ by fitting $\exp(St^2 \mu_3 a/r)$ to their experimental data, obtaining $\mu_3> 0$. However, we show in the Supplemental Material that the theory specifies $\mu_3\leq 0$, meaning this term inhibits clustering. Therefore, the scaling of the explosive RDF observation of Yavuz. \emph{et al.} \cite{yavuz18} cannot be justifiably associated with $\exp(St^2 \mu_3 a/r)$.

Finally, the leading order contribution in $St$ is $\exp(St \mu_2 a/r)$. In Yavuz \emph{et al} \cite{yavuz18} this contribution is absent since they argued that the third-order correlation on which $\mu_2$ depends is zero for isotropic turbulence. As a result of this they concluded that the leading-order effects of particle inertia occurs at $\mathcal{O} (St^2)$. However, as we demonstrate in the Supplemental Material, this correlation cannot be zero. It is associated with the average amplification of the strain-rate and vorticity fields in turbulence, which are associated with the energy cascade \cite{carbone20} and are necessarily finite in three-dimensional turbulence \cite{tsinober}. Not only does this mean that for the corrected theory, inertia affects $g(r)$ at $\mathcal{O}(St)$, instead of $\mathcal{O}(St^2)$, but also $\mu_2\leq 0$, so that the leading order effect of inertia in the presence of HI is to suppress the clustering, not enhance it. %Interestingly, for a two dimensional turbulent flow, the third order correlation is zero and so $\mu_2=0$, showing that the clustering of inertial particles due to HI depends upon the dimensionality of the flow.

\textbf{Experiments -} The experimental dataset presented in this paper was acquired by following the new small-$r$ particle tracking methodology developed in Hammond and Meng \cite{hammond21}. We measured $g(r)$ at 10 different $St$ and 4 different particle sizes, listed in Table \ref{Exp_table}. The result at $St=0.74, a=14.25\mu m$ is identical to those of \cite{hammond21}.

\begin{table}
	\centering
	\def\arraystretch{1.5}
	\begin{tabular}{cccccccc}
		\multicolumn{7}{l}{\textbf{Flow Conditions}}\\
		\multicolumn{2}{l}{Reynolds Number $Re_\lambda$}   & 246 & 277 & 324 & 334 & 357 \\
		\multicolumn{2}{l}{Kolmogorov Length $\eta$ ($\mu m$)}  & 179 & 141 & 123 & 109 & 101 \\
		\multicolumn{7}{l}{\textbf{Particle Properties}}\\
		\makecell{Radius \\ $a$ ($\mu m$) }  	& \makecell{Density \\ $\rho$ ($g/cm^3$) }&  \multicolumn{5}{c}{    \makecell{Stokes number\\ $St$ }      } \\
		$3.75 \pm 1.25$ & $0.95 \pm 0.05$ &  0.07 & 0.12 & 0.16 & 0.20 & 0.23 \\
		$8.75 \pm 1.25$  & $0.74 \pm 0.08$ & 0.23 & 0.37 &  &  &  \\
		$14.25 \pm 1.75$  & $0.31 \pm 0.02$ & 0.36 & 0.56 & 0.74 & 0.93 & 1.06 \\
		$20.75 \pm 1.75$  & $0.30 \pm 0.03$ & 0.74 & & & & \\
		\end{tabular}%
	\vspace{-1mm}
	\caption{Particle properties, Stokes numbers, and corresponding flow conditions in the experiments. For complete flow details see \cite{dou16}.}
	\label{Exp_table}
\end{table}

The experiments were performed in the enclosed, fan-driven, Homogeneous Isotropic Turbulence chamber (HIT chamber). The complete turbulence characteristics of this chamber are detailed in \cite{dou16}. The particles were hollow spheres by 3M (3M Glass Bubbles, types K25, S60, and IM16K), which allowed particle size control through sieving and inertia control through choice of particle type \cite{dou18b}. We sieved the originally widely polydisperse particles to acquire narrower, albeit still polydisperse, size distributions for each of the four particle samples. Particle density was measured with a Micromeritics accu-Pyc II 1340 gas pycnometer.

Uncertainties in $r$ and $g(r)$ were calculated following the method of \cite{hammond21} and are presented in the Supplemental Material. These two uncertainties had similar magnitudes across all conditions. Convergence of the RDF was achieved and the standard error was $<2\%$. For the $a=14.25\mu m$ particles, 15465 realizations were acquired for statistical convergence as in \cite{hammond21}. For the remaining three particle types, 9279 realizations were acquired.

% Hammond et al identified three important regimes in $r$ to isolate the relevant physics: the regime from large $r$ down to $r/\eta \approx 1$, corresponding to $r/a\approx O(10)$, known to be dominated by turbulence, the regime from $4<r/a<10$ where RDF grows explosively and scales as $r^{-6}$ due to particle-particle interactions, and the plateau regime in RDF from $2<r/a<4$ likely from polydispersity. In experiments, the RDF climbed from $\mathcal{O}(1)$ at $r/a\approx O(10)$ to $\mathcal{O}(1000)$ near to contact, far beyond prior theoretical predictions for this $r^{-6}$ behavior, and only roughly corroborated by the order of magnitude of the RDF near-contact in the experiments of Yavuz \emph{et al} \cite{yavuz18}.

\textbf{Results -} The experimental results for $g(r)$ all 13 flow and particle combinations are shown in Fig. \ref{RDF_plot}. At larger $r$, we observe the behavior $g(r)\sim r^{-St^2\mu_4}$ from \eqref{RDF_predic}. The RDF in this regime is consistent with previous experiments \cite{salazar08}. When $r/a$ decreases to $r/a\approx 30$ in Fig. \ref{RDF_plot}(a) for $a=3.75\mu m$ and $r/a\approx 12$ in Fig. \ref{RDF_plot}(b) for $a=14.25\mu m$, $g(r)$ grows explosively for all $St$, attaining values that are two orders of magnitude larger than those observed in previous simulations of inertial particles in turbulence without HI \cite{ireland16a}. In this explosive regime,  $g(r)-1\propto (r/a)^{-6}$. This is consistent with the far-field form of \eqref{RDF_predic} in the limit $St\to 0$.

When $r/a$ further decreases to below $r/a\approx 10$ in Fig. \ref{RDF_plot}(a) and $r/a\approx 3.5$ in Fig. \ref{RDF_plot}(b), $g(r)$ flattens out. This is most likely due to particle polydispersity, which is known to cause $g(r)$ to asymptote to a constant value at $r<r_c$, where $r_c$ is a cut-off scale that increases with increasing polydispersity in the system \cite{chun05,saw2012spatial1,saw2012spatial2,dhariwal18,bhatnagar18,momenifar19}. In our experiments, as particle size decreased, polydispersity increased (quantified by the ratio of the standard deviation to the mean of the particle size distribution), and correspondingly for smaller $a$ the flattened region in $g(r)$ broadened.
{\vspace{0mm}\begin{figure}[h]
		\centering
		\subfloat[]
		{\begin{overpic}
				[trim = 30mm 90mm 30mm 93mm,scale=0.48,clip,tics=20]{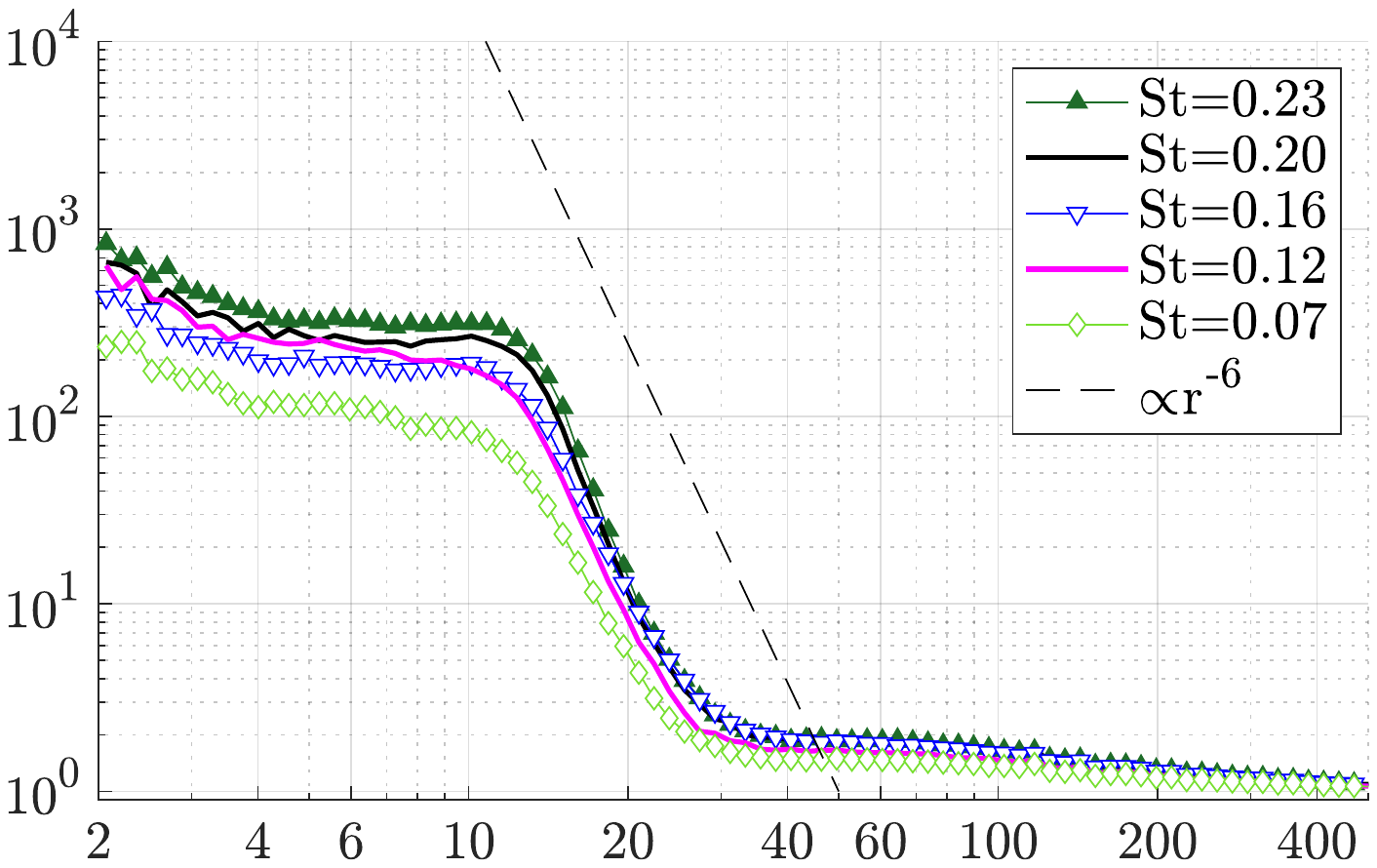}
				\put(105,0){$r/a$}
				\put(-5,64){\rotatebox{90}{$g(r)$}}
		\end{overpic}}\\
		\vspace{-4mm}\subfloat[]
		{\begin{overpic}
				[trim = 30mm 90mm 30mm 93mm,scale=0.48,clip,tics=20]{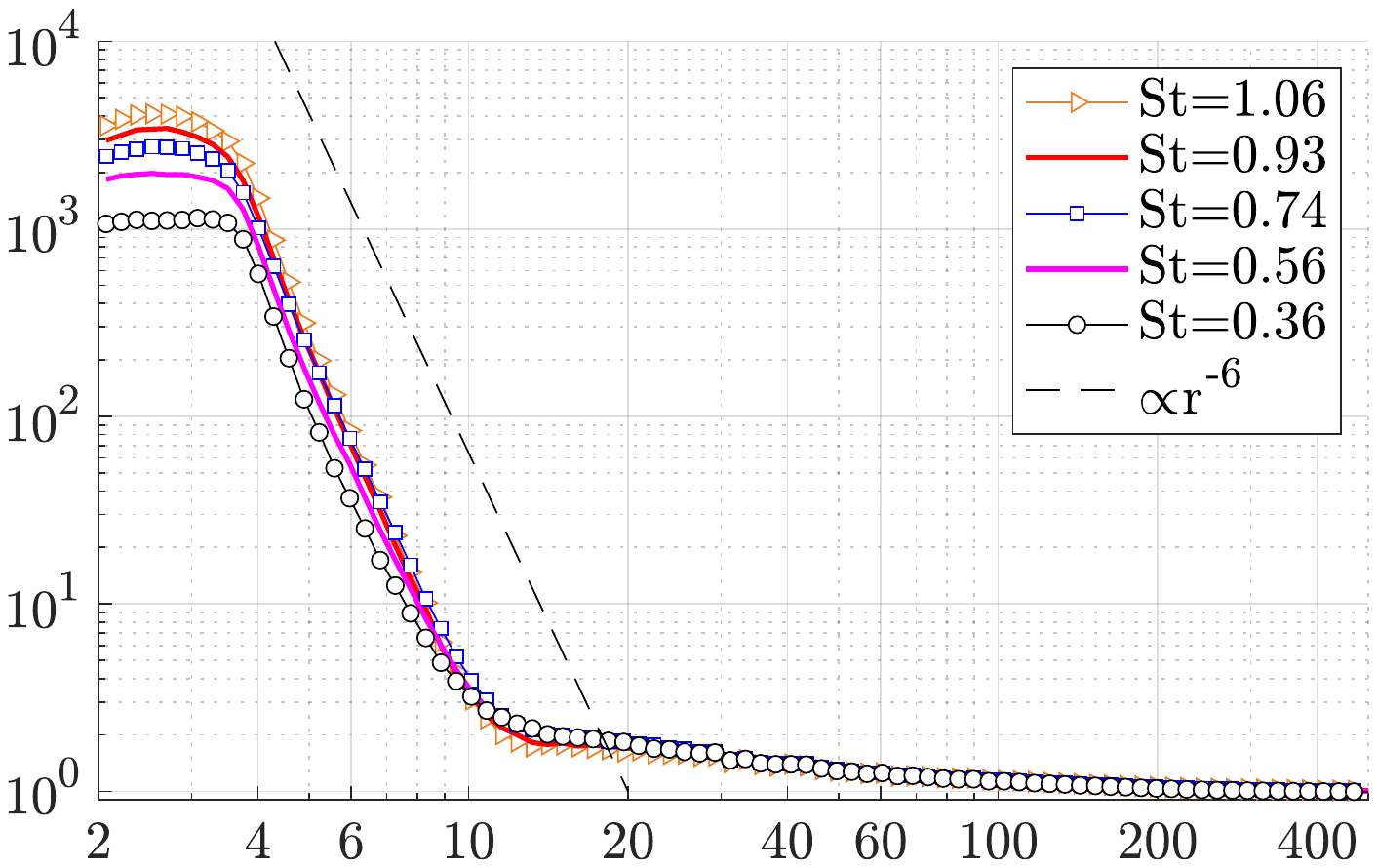}
				\put(105,0){$r/a$}
				\put(-5,64){\rotatebox{90}{$g(r)$}}
		\end{overpic}}\\
		\vspace{-4mm}\subfloat[]
		{\begin{overpic}
				[trim = 30mm 90mm 30mm 93mm,scale=0.48,clip,tics=20]{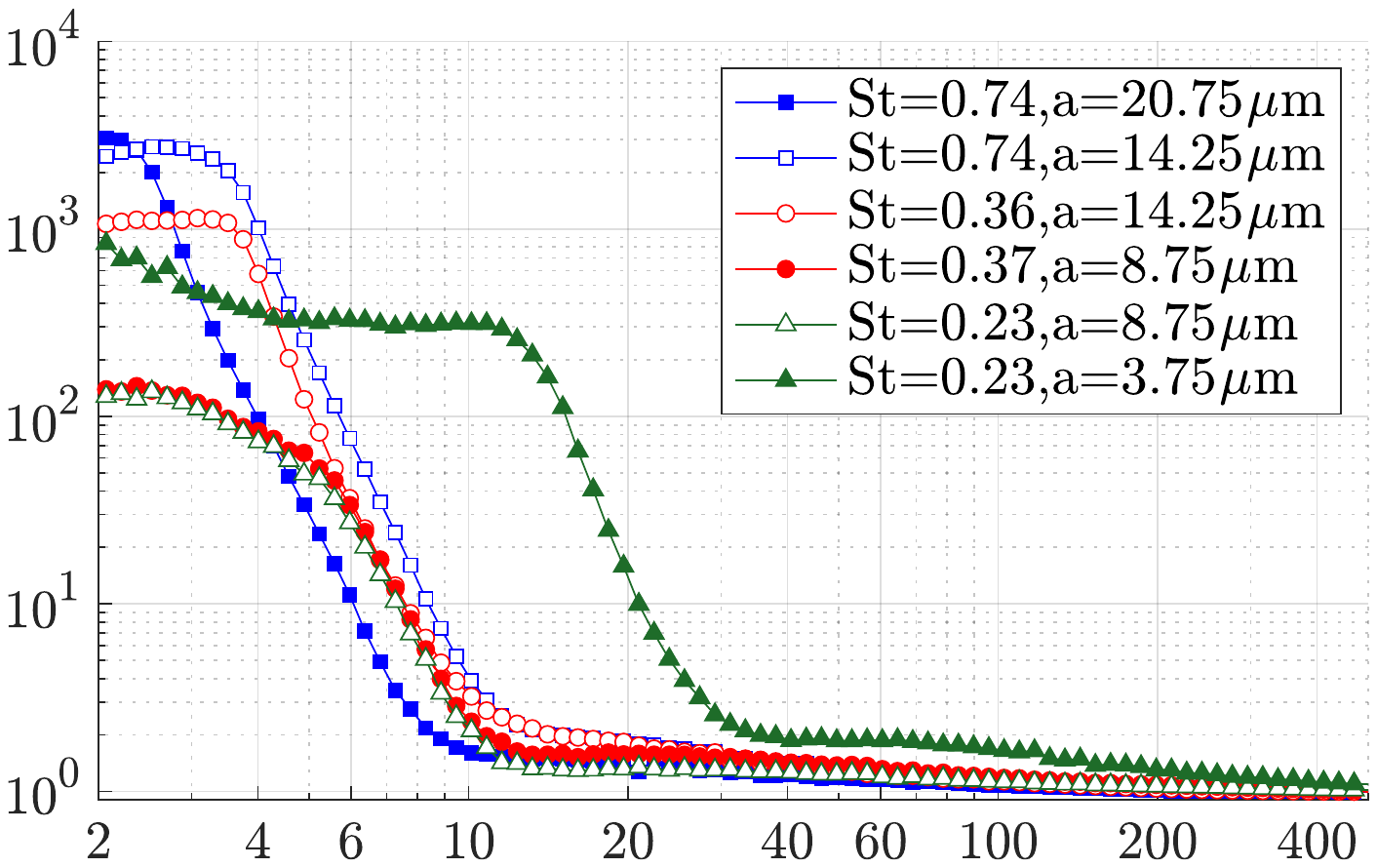}
				\put(105,0){$r/a$}
				\put(-5,64){\rotatebox{90}{$g(r)$}}
		\end{overpic}}\\
		\caption{RDF for different $St$ and particle radii. (a) $a=3.75\mu m$,  (b) $a=14.25\mu m$. Also shown is the behavior $g(r)\propto r^{-6}$. Plot (c) compares results with same/similar $St$ but different $a$ and $Re_\lambda$ (see Table \ref{Exp_table}).} 
		\label{RDF_plot}
\end{figure}}

Figure~\ref{RDF_plot}(a-b) also shows that for fixed $a$, in the explosive regime, $g(r)$ increases with increasing $St$, with the scaling $g(r)-1\propto (r/a)^{-6}$ preserved. This would indicate that increasing $St$ weakly but consistently enhances the RDF, within the uncertainty of the RDF measurement. However, even for the cases where $St\ll1$ such that the theory applies, this is fundamentally inconsistent with \eqref{RDF_predic}, according to which particle inertia does not affect the $r^{-6}$ scaling, and for the HI contributions, reduces rather than enhances the clustering.

To investigate if the RDF collapses on $r/a$ as predicted by the theory, in Fig.~\ref{RDF_plot} (c) we plot RDF at three different $St$, each obtained for two different particle radii $a$. It can be seen clearly that for $St=0.36$ and $St=0.37$ (red curves) the RDF generally collapses over decreasing $r$, up until $r/a\approx 6$. For the $St=0.74$ (blue) case, the results might collapse down to $r/a\approx 4$, as this is within the uncertainty of $r$ measurement. For the $St=0.23$ (green) case, the results clearly do not collapse, which is inconsistent with \eqref{RDF_predic}.

Using data from direct numerical simulations (DNS) to prescribe the fluid statistics on which $\mu_1$ depends (see Supplemental Material), we find $\mu_1\approx 31.98$. By contrast, for $St=0.07$ and $a=3.75\mu m$ we find the proportionality coefficient for the relation $g(r)-1\propto (r/a)^{-6}$ to be $3\times 10^8$, and for $St=0.23$ and $a=8.75\mu m$ it is $1.5\times 10^6$. This is an enormous discrepancy between the theory and experiments.

\textbf{Discussion -} Yavuz \emph{et al} \cite{yavuz18} claimed that their data does not allow them to observe $g(r)\sim \exp( \mu_1 a^6/r^6)$, but that they do observe $g(r)\sim \exp(St^2 \mu_3 a/r)$, to which they fit their data to indirectly obtain $\mu_3$. As discussed earlier, this claim is highly problematic because while their fit yields $\mu_3> 0$, the theory requires $\mu_3\leq 0$ (and using our DNS we estimate $\mu_3\approx -40.15$). As such, their observations cannot be justifiably associated with $g(r)\sim \exp(St^2 \mu_3 a/r)$. Moreover, for some of the cases, their fit to $g(r)\sim \exp(St^2 \mu_3 a/r)$ is not that strong. Indeed, their case with $a=10\mu m$, $St=0.19$ is quite well described by $g(r)-1\propto  (r/a)^{-6}$ over the range $6<r/a<11$, the same scaling we observe. However, just as found in our data, their data implies a proportionality constant orders of magnitude larger than the theoretical $\mu_1$.

All of these findings show that the extreme clustering observed here and in \cite{yavuz18} cannot be correctly described by a theory based on the HI of weakly inertial particle-pairs, contrary to the claims of \cite{yavuz18}. We have therefore sought to understand which assumptions in the theory may be responsible for its catastrophic failure to predict $g(r)$ (noting that although the theory correctly predicts the scaling $g(r)-1\propto (r/a)^{-6}$, it probably does so for the wrong reasons given the enormous quantitative errors). To this end, we investigated four potential error sources.

First, the theory assumes HI between a particle pair; however, many-body HI could occur if three or more particles are found within each other's hydrodynamic disturbance field. To test if many-body HI occurred in our experiments, we calculated the average number of particles in a sphere of radius $R$ around a test particle of radius $a$, conditioned on there being at least one satellite particle around the test particle, denoted by $\mathcal{N}(R\vert \theta)$. For particle-pairs, $\mathcal{N}(R\vert \theta)=1$, while $\mathcal{N}(R\vert \theta)\geq 2$ indicates more than two particles in the sphere. 

\begin{figure}[h]
		\centering		
		\includegraphics[trim = 103mm 68mm 0mm 52mm,scale=0.6,clip,tics=20]{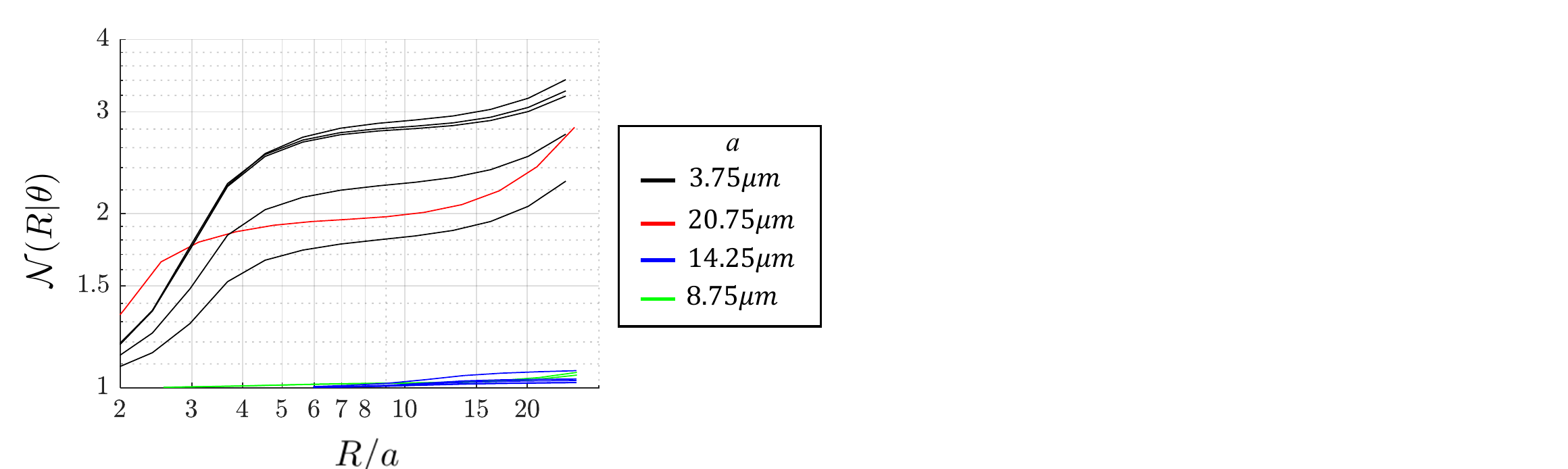}		
		\caption{Average number of particles (given that there are at least one of them) within a sphere of radius $R$ centered on a test particle, $\mathcal{N}(R\vert\theta)$, for all experiments.} 	
		\label{PinSph_plot}
\end{figure}

Figure~\ref{PinSph_plot} shows the results for $\mathcal{N}(R\vert \theta)$, where sphere size $R/a$ can be likened to separation $r/a$. We found that in the range of $r/a$ where $g(r)$ grows explosively the $14.25\mu m$ and $8.75 \mu m$ particles have $\mathcal{N}(R\vert \theta)$ close to 1, while the $3.75\mu m$ and $20.75\mu m$ particles have $\mathcal{N}(R\vert \theta)$ up to $3$. This variation is due to different particle number densities. Although this could mean that many-body HI is playing a role for the $3.75\mu m$ and $20.75\mu m$ particles, many-body HI definitely do not for the $14.25\mu m$ and $8.75 \mu m$ particles. Since extreme clustering is observed among all our experiments, many-body HI cannot be the fundamental cause of the discrepancy with the theory.

Second, particles were assumed smooth spheres in the theory. To test this assumption, we sampled particles from the flow facility during operation. Microscope images show that $87\%$ of particles were smooth spheres with no agglomeration (see Supplemental Material for images). Therefore this is not a cause of the discrepancy between theory and experiment.

Third, the neglect of other physically relevant forces on the particle-pair motion in the experiments, such as electrostatic and/or van der Waals forces etc. It is straightforward to show, however, that these forces would lead to behavior that is very different than $g(r)-1\propto  (r/a)^{-6}$ (see, e.g. \cite{lu10b}), and therefore cannot be the explanation. Moreover, we have measured the charge level in the flow facility, determining it as $<\mathcal{O}(10^{-16}) C$ (see Supplemental Material). Therefore, Coulomb forces on the particles were negligible.

Fourth, the particle Reynolds number $Re_p$ was assumed to be small in the theory \cite{bkl97,chun05,yavuz18} such that Stokes flow around the particles is assumed. If we use the expression $Re_p= a^2/\tau_\eta\nu$ \cite{bkl97} then for our experiments, $Re_p\ll 1$. Therefore, this assumption holds.
 
\textbf{Conclusions -} More experimental evidence of extreme clustering of inertial particles at small separations in a turbulent flow corroborates earlier observations \cite{yavuz18,hammond21} and allows for a clearer look into the scaling of $g(r)$ and the influence of $St$ and $a$. Our data confirms $g(r)-1\propto r^{-6}$ in the explosive scaling regime, contrary to Yavuz \emph{et al} \cite{yavuz18}. We demonstrate that the corrected theory based on weakly inertial particle-pair HI cannot explain the extreme clustering, since the theory predicts an inhibition rather than enhancement of $g(r)$ by the inertial contribution to HI, while in experiments St weakly increases the extreme clustering. Moreover, the theoretical predictions for the RDF are in error by orders of magnitude. As such, the extreme clustering observed here and in \cite{yavuz18} remains something of a mystery. The particle equation of motion invoked in the theory is clearly missing some vital effect, which future work must seek to uncover.

\bibliographystyle{unsrt}
\bibliography{Paper}

\end{document}

\maketitle

\section{Theory}

\subsection{Expression for the RDF}

Let $\bm{r}^p(t)$ denote the separation of a particle-pair, $\bm{w}^p(t)\equiv\dot{\bm{r}}^p(t)$ their relative velocity, and $\bm{r}$ a time-independent coordinate field. The exact solution for the probability density function (PDF) $\rho(\bm{r},t)\equiv\langle \delta(\bm{r}^p(t)-\bm{r})\rangle$ is \cite{tom19}
\begin{align}
\rho(\bm{r},t)=V^{-1}\Bigg\langle\exp\Bigg(-\int^t_0\bm{\nabla\cdot \mathcal{W}}(\bm{\xi}(s),s)\,ds\Bigg) \Bigg\rangle_{\bm{r}},\label{rhosol}
\end{align}
where we have assumed the particles are initially uniformly distributed over the domain $V$, $\langle\cdot\rangle_{\bm{r}}$ denotes an ensemble average conditioned on $\bm{\xi}(t)=\bm{r}$, where $\bm{\xi}$ is defined by $\partial_s\bm{\xi}\equiv\bm{\mathcal{W}}$. The formal definition of the field $\bm{\mathcal{W}}$ is
\begin{align}
\bm{\mathcal{W}}(\bm{r},t)\equiv \Big\langle\bm{w}^p(t)\Big\rangle^{\bm{r}^p(0),\bm{w}^p(0)}_{\bm{r},\bm{u}},\label{Field_def}
\end{align}
where the operator $\langle\cdot\rangle^{\bm{r}^p(0),\bm{w}^p(0)}_{\bm{r},\bm{u}}$ denotes an ensemble average over all initial separations and initial relative velocities, conditioned on a given realization of the flow $\bm{u}$, and conditioned on $\bm{r}^p(t)=\bm{r}$. This field reduces to $\bm{\mathcal{W}}(\bm{r},t)= \bm{w}^p(t\vert\bm{r})$ in the limit where there is a unique initial condition that generates a trajectory satisfying $\bm{r}^p(t)=\bm{r}$, such as is the case for fluid particles.

For a statistically stationary, isotropic system, the dependence upon $\bm{r}$ reduces to a dependence upon $r\equiv\|\bm{r}\|$, and the PDF $\rho(r)$ is related to the radial distribution function (RDF) as
\begin{align}
g(r)=\frac{N(N-1)}{n^2V}\rho(r),\label{RDF_rho}
\end{align}
where $N$ is the total number of particles in the flow, and $n\equiv N/V$. In the thermodynamic limit, $g(r)=V\rho(r)$ and we therefore finally obtain
\begin{align}
g(r)=\Bigg\langle\exp\Bigg(-\int^t_0\bm{\nabla\cdot \mathcal{W}}(\bm{\xi}(s),s)\,ds\Bigg) \Bigg\rangle_{r},\label{gsol}
\end{align}
which is the result quoted in the paper.

\subsection{Drift-diffusion model for the RDF}

In \cite{yavuz18}, the authors developed a theoretical model for how weakly inertial particles cluster in the presence of HI and turbulence. The basis for their result was the drift-diffusion models of \cite{bkl97,chun05}. However in going through their analysis we have found several significant errors. Therefore, in the following, we re-derive the correct form of the theory and discuss the differences with the result obtained in \cite{yavuz18}.

Following \cite{chun05}, the steady-state form of $\rho$ is determined by the transport equation (it is to be understood that $t\to \infty$)
\begin{align}
\frac{\partial}{\partial\bm{r}}\bm{\cdot}\Big(\bm{q}^{D}+ \bm{q}^{d}\Big)&={0}\label{rho_eq}\\
\bm{q}^{D}(\bm{r},t)&\equiv -\int_0^t\Bigg\langle\bm{\mathcal{W}}(\bm{r},t)\bm{\mathcal{W}}(\bm{\xi}(s),s)\bm{\cdot\nabla}\rho(\bm{\xi}(s))\Bigg\rangle_{\bm{r}}\,ds,\\
\bm{q}^{d}(\bm{r},t)&\equiv-\int_0^t \Bigg\langle\bm{\mathcal{W}}(\bm{r},t)\bm{\nabla\cdot \mathcal{W}}(\bm{\xi}(s),s)\rho(\bm{\xi}(s)) \Bigg\rangle_{\bm{r}}\,ds,
\end{align}
where $\bm{q}^{D}$ and $\bm{q}^{d}$ are the diffusion and drift vectors, respectively.
%It is important to appreciate that while not clearly stated in \cite{chun05}, the field $\bm{\mathcal{W}}$ is in fact not only a function of the particle pair separation but also the position of the primary particle $\bm{x}^p(t)$. While this coordinate dependence need not be explicitly stated since we are considering a flow that is statistically homogeneous in space, this dependence must be born in mind when we go on to construct expressions for $\bm{\mathcal{W}}$.

The transport equation above provides an integro-differential equation for the PDF $\rho(\bm{r})$. In the limit where the correlation timescale of the flow is very small, the equation becomes differential, since over these short correlation times $\bm{\xi}(s)\approx \bm{r}$, and this is referred to as the ``diffusion approximation'' \cite{bkl97,chun05}. As discussed in \cite{chun05}, the diffusion approximation is not valid in real turbulence since the correlation time of the flow is of the same order as that on which $\bm{\xi}(s)$ evolves. Nevertheless, in \cite{chun05} it is demonstrated that in real turbulence, the diffusion approximation gives the correct functional forms in the model, and the quantitative error associated with the diffusion approximation can be corrected for using a ``non-local'' correction coefficient, for which a model was constructed in \cite{chun05}. In view of this, we will also adopt the diffusion approximation (with the non-local correction coefficient to be explicitly included later)
\begin{align}
\bm{q}^{D}(\bm{r},t)&\approx -\int_0^t\Big\langle\bm{\mathcal{W}}(\bm{r},t)\bm{\mathcal{W}}(\bm{r},s)\Big\rangle_{\bm{r}}\,ds \bm{\cdot\nabla}\rho(\bm{r})\,\\
\bm{q}^{d}(\bm{r},t)&\approx-\rho(\bm{r})\int_0^t \Big\langle\bm{\mathcal{W}}(\bm{r},t)\bm{\nabla\cdot \mathcal{W}}(\bm{r},s) \Big\rangle_{\bm{r}}\,ds,
\end{align}
which is the form of the drift-diffusion model used in \cite{yavuz18}. To construct solutions for $\rho(\bm{r})$ in the regime $St\ll1$, we must specify the correlations in $\bm{q}^{D}$ and $\bm{q}^{d}$, which are constructed using solutions to the particle equation of motion.

Just as in \cite{yavuz18}, we consider monodisperse pairs of small, heavy, inertial particles, subject to HI, whose equation of motion is 
\begin{align}
\bm{w}^p(t)=St\tau_\eta\mathbf{J}^p\bm{\cdot}\dot{\bm{w}}^p+\bm{f}^p+St\tau_\eta\frac{3a}{5\|\bm{r}^p\|}C\bm{\dot{\Upsilon}}^p\times\bm{ r}^p,\label{reom}
\end{align}
where $\bm{\Upsilon}^p$ is the total angular velocity of the two particles (sum of their angular velocities), $\mathbf{J}^p=\mathbf{J}(\bm{x}^p(t),\bm{r}^p(t),t)$, $\bm{f}^p=\bm{f}(\bm{x}^p(t),\bm{r}^p(t),t)$,
\begin{align}
\mathbf{J}(\bm{x},\bm{r},t)&\equiv\Bigg(A\frac{\bm{r}\bm{r}}{\|\bm{r}\|^2} +B\Bigg[\mathbf{I}-\frac{\bm{r}\bm{r}}{\|\bm{r}\|^2}\Bigg] \Bigg),\\
\bm{f}(\bm{x},\bm{r},t)&\equiv\bm{\Gamma}\bm{\cdot r}-2a\Bigg(D\frac{\bm{r}\bm{r}}{\|\bm{r}\|^2} +E\Bigg[\mathbf{I}-\frac{\bm{r}\bm{r}}{\|\bm{r}\|^2}\Bigg] \Bigg)\bm{\cdot}\frac{(\bm{S}\bm{\cdot r})}{\|\bm{r}\|},
\end{align}
and $\bm{x}$ refers to the arguments of the velocity gradient $\bm{\Gamma}(\bm{x},t)\equiv \bm{\nabla u}$ and the strain-rate $\bm{S}(\bm{x},t)\equiv( \bm{\nabla u}+ \bm{\nabla u}^\top)/2$. The terms $A,B,C,D,E$ are nondimensional functions of $r\equiv\|\bm{r}\|$ only, and we will discuss them in more detail momentarily.

Note that in the expression for $\bm{f}$ above, it has been assumed that $\|\bm{r}\|$ is sufficiently small so that the background flow (i.e. the flow field in the absence of HI) is locally linear. In a turbulent flow this formally requires assuming $\|\bm{r}\|\ll \eta$, where $\eta$ is the Kolmogorov length scale. However, it should be appreciated that in practice this locally linear flow assumption is known to be valid up to $\|\bm{r}\|=O(10\eta)$ (e.g. see figure 12 of \cite{ireland16a}). This is due to the fact that the Kolmogorov theory underestimates the scale at which viscous and inertial forces balance, which may be due in part to the phenomena of nonlinear depletion \cite{frisch}.

By simply inserting the expression for $\bm{w}^p(t)$ given in \eqref{reom} into the term $\dot{\bm{w}}^p$ that appears in \eqref{reom}, then to order $\mathcal{O}(St^2)$ we obtain
\begin{align}
\bm{w}^p(t)=St\tau_\eta\mathbf{J}^p\bm{\cdot}\frac{d}{dt}\bm{f}^p+\bm{f}^p+St\tau_\eta\frac{3a}{5\|\bm{r}^p\|}C\bm{\dot{\Upsilon}}^p\times\bm{ r}^p+\mathcal{O}(St^2).
\end{align}
%
%where $\bm{\Omega}^p$ is the fluid vorticity at the primary particle position. 

Then, based on the definition of the field $\bm{\mathcal{W}}(\bm{r},t)$ given in \eqref{Field_def} we obtain
\begin{align}
{\bm{\mathcal{W}}}=St\tau_\eta\mathbf{J}\bm{\cdot}\frac{D}{Dt}\bm{f}+\bm{f}+St\tau_\eta\frac{3a}{5 r}C\dot{\bm{\Theta}}\times\bm{ r}+\mathcal{O}(St^2),\label{WlowSt}
\end{align}
where
\begin{align}
\dot{\bm{{\Theta}}}(\bm{r},t)\equiv \Big\langle\bm{\dot{\Upsilon}}^p\Big\rangle^{\bm{r}^p(0),\bm{w}^p(0)}_{\bm{r},\bm{u}}.
\end{align}
Note that owing to the definition of $\bm{\mathcal{W}}(\bm{r},t)$, in \eqref{WlowSt} the terms $\mathbf{J}$ and $\bm{f}$ are explicitly $\mathbf{J}(\bm{x}^p(t),\bm{r},t)$ and $\bm{f}(\bm{x}^p(t),\bm{r},t)$, i.e. measured at fixed separation $\bm{r}$, but along the time-dependent trajectory $\bm{x}^p(t)$.

For clarity, we now switch to index notation, and write the result for ${\bm{\mathcal{W}}}$ in the form of an expansion in $St$
\begin{align}
\mathcal{W}_i&= \mathcal{W}_i^{[0]}+St \mathcal{W}_i^{[1]}+\mathcal{O}(St^2),\\
\mathcal{W}_i^{[0]}&\equiv f_i,\\
\mathcal{W}_i^{[1]}&\equiv \tau_\eta J_{ij} \frac{D}{Dt}f_j+\tau_\eta\frac{3a}{5 r}C\epsilon_{ijk}\dot{\Theta}_j r_k,
\end{align}
so that to order $St^2$ we have
\begin{align}
\begin{split}
{q}_i^d(\bm{r},t)=&-\rho(\bm{r})\int_0^t \Bigg\langle\mathcal{W}^{[0]}_i(\bm{r},t)\frac{\partial}{\partial r_l}{\mathcal{W}}^{[0]}_l(\bm{r},s)\,ds\Bigg\rangle\\
&-St \rho(\bm{r})\int_0^t \Bigg\langle\mathcal{W}^{[0]}_i(\bm{r},t)\frac{\partial}{\partial r_l}{\mathcal{W}}^{[1]}_l(\bm{r},s)\,ds\Bigg\rangle\\
&-St \rho(\bm{r})\int_0^t \Bigg\langle\mathcal{W}^{[1]}_i(\bm{r},t)\frac{\partial}{\partial r_l}{\mathcal{W}}^{[0]}_l(\bm{r},s)\,ds\Bigg\rangle\\
&-St^2 \rho(\bm{r})\int_0^t\Bigg\langle \mathcal{W}^{[1]}_i(\bm{r},t)\frac{\partial}{\partial r_l}{\mathcal{W}}^{[1]}_l(\bm{r},s)\,ds\Bigg\rangle.
\end{split}
\label{drift_expansion}
\end{align}
Following \cite{yavuz18}, we will focus on the far-field asymptotic behavior where $a/r\ll 1$, retaining at each order in $St$ only the leading order contributions from HI. In order to obtain these, we must use the far-field forms of $\bm{f}$ and $\mathbf{J}$, which are obtained using the far-field asymptotic relations $A(r)\sim -1+3a/2r$, $B(r)\sim -1+3a/4r$, $D(r)\sim 5a^2/2 r^2 - 4 a^4/r^4 + 25 a^5/2 r^5$, $E(r)\sim 8 a^4/3r^4$ \cite{batchelor72,kim}. 

The required far-field forms are then
\begin{align}
\mathcal{W}_i^{[0]}&\sim \Gamma^p_{ij}r_j,\\
\frac{\partial}{\partial r_l}{\mathcal{W}}^{[0]}_l(\bm{r},t)&\sim 75\frac{a^6}{r^6}S^p_{kn}\frac{r_k r_n}{r^2},\\
\mathcal{W}^{[1]}_i(\bm{r},t)&\sim \tau_\eta\Big(\frac{3a}{4r}-1\Big)\Gamma_{ik}^p\Gamma^p_{km}r_m+\tau_\eta \frac{3a}{4r}\frac{r_i r_j r_m}{r^2} \Gamma_{jk}^p\Gamma^p_{km},\\
\frac{\partial}{\partial r_l}{\mathcal{W}}^{[1]}_l(\bm{r},t)&\sim\tau_\eta \Big(\frac{3a}{4r}-1\Big)\Gamma^p_{nm}\Gamma^p_{mn}+\tau_\eta\frac{3a}{4r}\frac{r_j r_n}{r^2}\Gamma^p_{jm}\Gamma^p_{mn}.
\end{align}
Here we have thrown away the terms involving $\dot{\bm{{\Theta}}}$, since as noted in \cite{yavuz18}, this term is sub-leading in the far-field under the assumptions $St\ll1$ and that the local flow around the particles is steady Stokes flow (already assumed in the particle equation of motion). The results above match exactly with those in \cite{yavuz18} (although our notation differs), and retain the leading order contributions from the HI effects.

The first contribution in \eqref{drift_expansion} is 

\begin{align}
\begin{split}
\int_0^t \Bigg\langle\mathcal{W}^{[0]}_i(\bm{r},t)\frac{\partial}{\partial r_l}{\mathcal{W}}^{[0]}_l(\bm{r},s)\,ds\Bigg\rangle&\sim 75\frac{a^6}{r^6}\frac{r_j  r_m r_n}{r^2}\int_0^t \Big\langle  \Gamma^p_{ij}(t) S^p_{mn}(s)\Big\rangle\,ds\\
&=75\frac{a^6}{r^6}\frac{r_j  r_m r_n}{r^2}   \frac{1}{10}\langle S^p_{ab}(t)S^p_{ab}(t)\rangle\Big(\delta_{im}\delta_{jn}+\delta_{in}\delta_{jm}-(2/3)\delta_{ij}\delta_{mn} \Big)\tau_S  \\
&=10\frac{a^6}{r^6}\tau_S  r_i \langle  S^2\rangle,
\end{split}
\end{align}
where $\tau_S$ is the correlation timescale of the strain-rate, and $\langle  S^2\rangle\equiv  \langle S^p_{ab}(t)S^p_{ab}(t)\rangle$. This result is the same as the far-field version of the drift velocity derived in \cite{bkl97}.

In \cite{yavuz18}, it is argued that the order $St$ contributions to $\bm{q}^d$ disappear due to isotropy of the flow. We now investigate this claim. The second contribution in \eqref{drift_expansion} involves
\begin{align}
\begin{split}
\int_0^t \Bigg\langle\mathcal{W}^{[0]}_i(\bm{r},t)\frac{\partial}{\partial r_l}{\mathcal{W}}^{[1]}_l(\bm{r},s)\,ds\Bigg\rangle\sim\tau_\eta  \Big(\frac{3a}{4r}-1\Big)r_j \int_0^t \Big\langle\Gamma^p_{ij}(t)
\Gamma^p_{nm}(s)\Gamma^p_{mn}(s)\Big\rangle\,ds\\
+\tau_\eta \frac{3a}{4r}\frac{r_j r_k r_n}{r^2} \int_0^t \Big\langle\Gamma^p_{ij}(t)\Gamma^p_{km}(s)\Gamma^p_{mn}(s)\Big\rangle\,ds.
\end{split}
\end{align}
Invoking isotropy, incompressibility and the Betchov relations \cite{betchov56} we have
\begin{align}
\Big\langle\Gamma^p_{ij}(t)\Gamma^p_{nm}(t)\Gamma^p_{mn}(t)\Big\rangle=\frac{\delta_{ij}}{3}\Big\langle\Gamma^p_{aa}(t)\Gamma^p_{nm}(s)\Gamma^p_{mn}(s) \Big\rangle=0,
\end{align}
and 
\begin{align}
\begin{split}
\Big\langle\Gamma^p_{ij}(t)\Gamma^p_{km}(s)\Gamma^p_{mn}(s)\Big\rangle=\frac{1}{30}
\Big\langle\Gamma^p_{bc}(t)\Gamma^p_{bd}(s)\Gamma^p_{dc}(s)\Big\rangle\Big(4\delta_{ik}\delta_{jn}-\delta_{ij}\delta_{kn} -\delta_{in}\delta_{jk} \Big).
\end{split}
\end{align}
In \cite{yavuz18}, it is assumed that this quantity is zero, but we will now show that it is not. Just as in \cite{chun05,yavuz18} we may write this as
\begin{align}
\Big\langle\Gamma^p_{bc}(t)\Gamma^p_{bd}(s)\Gamma^p_{dc}(s)\Big\rangle=\Big\langle\Gamma^p_{bc}(t)\Gamma^p_{bd}(t)\Gamma^p_{dc}(t)\Big\rangle\Sigma(t,s),
\end{align}
where $\Sigma$ denotes the autocorrelation for the quantity. By introducing the strain-rate and vorticity, the single-time invariant can be written as
\begin{align}
\Big\langle\Gamma^p_{bc}(t)\Gamma^p_{bd}(t)\Gamma^p_{dc}(t)\Big\rangle=\Big\langle{S}^p_{bc}(t){S}^p_{bd}(t){S}^p_{dc}(t)\Big\rangle-\frac{1}{4}\Big\langle{S}^p_{bc}(t)\Omega^p_{b}(t)\Omega^p_{c}(t)\Big\rangle.
\end{align}
This invariant is not zero; the first term on the rhs is the strain-rate production term, and the second is the enstrophy production term (i.e. vortex stretching) \cite{tsinober}. These invariants are always finite in three dimensional turbulence, and are directly connected to the energy cascade process \cite{carbone20}. Thus, remarkably, the processes governing the energy cascade also contribute to inertial particle clustering in the presence of HI. Note, however, that these terms are zero for two dimensional turbulence.

Putting this together we obtain

\begin{align}
\begin{split}
\int_0^t \Bigg\langle\mathcal{W}^{[0]}_i(\bm{r},t)\frac{\partial}{\partial r_l}{\mathcal{W}}^{[1]}_l(\bm{r},s)\,ds\Bigg\rangle\sim\frac{a \tau_\eta \tau_{\Sigma}}{20 r}r_i\langle \Gamma^3\rangle,
\end{split}
\end{align}
where $\tau_{\Sigma}$ is the timescale associated with the autocorrelation $\Sigma$, and $\langle \Gamma^3\rangle\equiv \langle\Gamma^p_{bc}(t)\Gamma^p_{bd}(t)\Gamma^p_{dc}(t)\Big\rangle$. Note that since $\langle \Gamma^3\rangle<0$ in three-dimensional turbulence \cite{tsinober}, then this contribution actually \emph{opposes} the clustering of the particles.
 
The third contribution in \eqref{drift_expansion}  
\begin{align}
\begin{split}
\int_0^t\Bigg\langle \mathcal{W}^{[1]}_i(\bm{r},t)\frac{\partial}{\partial r_l}{\mathcal{W}}^{[0]}_l(\bm{r},s)\,ds\Bigg\rangle=&75\frac{ a^6}{r^6}\tau_\eta\frac{r_m r_l r_q}{r^2}\Big(\frac{3a}{4r}-1\Big)\int_0^t \Big\langle\Gamma^p_{ik}(t)\Gamma^p_{km}(t)S^p_{lq}(s)\Big\rangle\,ds\\
&+\frac{225}{4}\frac{ a^7}{r^7}\tau_\eta\frac{r_i r_j r_m r_l r_q}{r^4}\int_0^t \Big\langle\Gamma^p_{jk}(t)\Gamma^p_{km}(t)S^p_{lq}(s)\Big\rangle\,ds
\end{split}
\end{align}
Similar to before 
\begin{align}
\begin{split}
\Big\langle\Gamma^p_{ik}(t)\Gamma^p_{km}(t)S^p_{lq}(s)\Big\rangle=\frac{1}{60}
\Big\langle\Gamma^p_{bc}(t)\Gamma^p_{cd}(t)\Gamma^p_{bd}(t)\Big\rangle\Big(4\delta_{il}\delta_{mq}-\delta_{im}\delta_{lq} -\delta_{iq}\delta_{ml} \Big)
\Sigma'(t,s),
\end{split}
\end{align}
where $\Sigma'$ denotes the autocorrelation for the quantity, and we have used the Betchov relations to obtain
\begin{align}
\begin{split}
\Big\langle\Gamma^p_{bc}(t)\Gamma^p_{cd}(t)S^p_{bd}(t)\Big\rangle=\frac{1}{2}\Big\langle\Gamma^p_{bc}(t)\Gamma^p_{cd}(t)\Gamma^p_{bd}(t)\Big\rangle,
\end{split}
\end{align}
Putting this altogether we obtain
\begin{align}
\begin{split}
\int_0^t \Bigg\langle\mathcal{W}^{[1]}_i(\bm{r},t)\frac{\partial}{\partial r_l}{\mathcal{W}}^{[0]}_l(\bm{r},s)\Bigg\rangle\,ds&=\frac{5}{2}\Big(\frac{3 a}{2r}-1   \Big)\frac{a^6}{r^6}\tau_\eta\tau_{\Sigma'} r_i\langle \Gamma^3\rangle,
\end{split}
\end{align}
 where $\tau_{\Sigma'}$ is the timescale associated with the autocorrelation $\Sigma'$.

The fourth contribution in \eqref{drift_expansion}  

\begin{align}
\begin{split}
&\int_0^t \Bigg\langle\mathcal{W}^{[1]}_i(\bm{r},t)\frac{\partial}{\partial r_l}{\mathcal{W}}^{[1]}_l(\bm{r},s)\Bigg\rangle\,ds\\
&=  \tau_\eta^2\Big(\frac{3a}{4 r}-1\Big)^2 r_m 
\int_0^t \Bigg\langle\Gamma^p_{ik}(t)\Gamma^p_{km}(t)\Gamma^p_{np}(s)\Gamma^p_{pn}(s)   \Bigg\rangle\,ds\\
&+\tau_\eta^2\frac{3a}{4 r}\Big(\frac{3a}{4 r}-1\Big) \frac{r_m r_j r_n}{r^2} \int_0^t \Bigg\langle\Gamma^p_{ik}(t)\Gamma^p_{km}(t)\Gamma^p_{jp}(s)\Gamma^p_{pn}(s)   \Bigg\rangle\,ds\\
&+\tau_\eta^2\frac{3a}{4 r}\Big(\frac{3a}{4 r}-1\Big) \frac{r_i r_j r_m}{r^2} \int_0^t \Bigg\langle\Gamma^p_{jk}(t)\Gamma^p_{km}(t)\Gamma^p_{np}(s)\Gamma^p_{pn}(s)   \Bigg\rangle\,ds\\
&+\tau_\eta^2\Big(\frac{3a}{4 r}\Big)^2 \frac{r_i r_j r_m r_p r_q}{r^4} \int_0^t \Bigg\langle\Gamma^p_{jk}(t)\Gamma^p_{km}(t)\Gamma^p_{pl}(s)\Gamma^p_{lq}(s)   \Bigg\rangle\,ds
\end{split}
\end{align}
These integrals involve the quantity (with differing index labels)
\begin{align}
A_{imjn}(t,s)&\equiv\Big\langle\Gamma^p_{ik}(t)\Gamma^p_{km}(t)\Gamma^p_{jp}(s)\Gamma^p_{pn}(s)   \Big\rangle,
\end{align}
and this may be re-written using isotropy and an autocorrelation function as
\begin{align}
A_{imjn}(t,s)&=\frac{\langle \Gamma^4_1\rangle}{30}\Big(4\delta_{im}\delta_{jn}-\delta_{ij}\delta_{mn}-\delta_{in}\delta_{mj}\Big)\Psi_1(t,s)\\
&+\frac{\langle \Gamma^4_2\rangle}{30}\Big(-\delta_{im}\delta_{jn}+4\delta_{ij}\delta_{mn}-\delta_{in}\delta_{mj}\Big)\Psi_2(t,s)\\
&+\frac{\langle \Gamma^4_2\rangle}{30}\Big(-\delta_{im}\delta_{jn}-\delta_{ij}\delta_{mn}+4\delta_{in}\delta_{mj}\Big)\Psi_3(t,s).
\end{align}
where $\langle \Gamma^4_1\rangle\equiv A_{aabb}(t,t)$,  $\langle \Gamma^4_2\rangle\equiv A_{abab}(t,t)$, $\langle \Gamma^4_3\rangle\equiv A_{abba}(t,t)$. Using this we then obtain for the fourth contribution in \eqref{drift_expansion}  
\begin{align}
\begin{split}
&\int_0^t \Bigg\langle\mathcal{W}^{[1]}_i(\bm{r}^p(t),t)\frac{\partial}{\partial r_l}{\mathcal{W}}^{[1]}_l(\bm{r}^p(s),s)\Bigg\rangle\,ds\\
&=  \tau_\eta^2\tau_{\Psi_1}\frac{\langle \Gamma^4_1\rangle}{3}r_i\Big(\frac{3a}{4 r}-1\Big)\Big(\frac{3a}{2r}-1\Big)\\
&+\frac{2}{15}\Big(\Big(\frac{3a}{4 r}\Big)^2-\frac{3a}{8 r}\Big)\tau_\eta^2r_i\Big(\langle \Gamma^4_1\rangle\tau_{\Psi_1}+\langle \Gamma^4_2\rangle\tau_{\Psi_2}+\langle \Gamma^4_3\rangle\tau_{\Psi_3} \Big).
\end{split}
\end{align}
We now gather together all the contributions to \eqref{drift_expansion}, and at each order in $St$ only retain up to the leading order contribution from HI, yielding 
\begin{align}
\begin{split}
q_i^d=&-\rho10\frac{a^6}{r^6}\tau_S\langle S^2\rangle r_i
-\rho St\tau_\eta\tau_{\Sigma}\frac{a}{20r}\langle \Gamma^3\rangle r_i
-\rho St^2\tau_\eta^2 \tau_{\Psi_1}\Big(1-\frac{9a}{4r} \Big)\frac{\langle \Gamma^4_1\rangle}{3}r_i\\
&+\rho St^2\tau_\eta^2\frac{a}{20r}\Big(\langle \Gamma^4_1\rangle\tau_{\Psi_1}+\langle \Gamma^4_2\rangle\tau_{\Psi_2}+\langle \Gamma^4_3\rangle\tau_{\Psi_3} \Big)r_i.
\end{split}\label{drift}
\end{align}

Concerning the diffusion term, in the far-field this is to leading order \cite{yavuz18}
\begin{align}
\begin{split}
q^D_{i}(\bm{r})\sim-r_i B^{nl}r\tau_\eta^{-1} \nabla\rho,\label{Diff}
\end{split}
\end{align}
where $B_{nl}$ is the non-local correction coefficient discussed earlier. Using \eqref{Diff} and \eqref{drift} in \eqref{rho_eq}, together with \eqref{RDF_rho}, we obtain the following solution for the RDF

\begin{align}
\begin{split}
g(r)\sim \exp\Bigg(\mu_1\frac{a^6}{r^6}\Bigg)\exp\Bigg( (St\mu_2 +St^2\mu_3)\frac{a}{r}\Bigg)\Big(\frac{r}{a}\Big)^{-St^2\mu_4},\label{RDF_sol}
\end{split}
\end{align}
where 

\begin{align}
\mu_1&\equiv 5\tau_\eta\tau_S\langle S^2\rangle/3B^{nl},\\
\mu_2&\equiv \tau_\eta^2\tau_\Sigma\langle\Gamma^3\rangle/20B^{nl},\\
\mu_3 &\equiv  -9\tau_\eta^3\tau_{\Psi_1}\langle \Gamma^4_1\rangle/12 B^{nl}-\tau_\eta^3\Big(\langle \Gamma^4_1\rangle\tau_{\Psi_1}+\langle \Gamma^4_2\rangle\tau_{\Psi_2}+\langle \Gamma^4_3\rangle\tau_{\Psi_3} \Big)/20 B^{nl},\\
\mu_4&\equiv  \tau_\eta^3 \tau_{\Psi_1}\langle \Gamma^4_1\rangle/3 B^{nl}.
\end{align}
Note that we have used $\sim$ in the solution for $g(r)$ to reflect the fact that for $St\ll 1$ there is an $\mathcal{O}(1)$ boundary condition at $r=\eta$ \cite{chun05,ireland16a}.

This result in \eqref{RDF_sol} differs from that of \cite{yavuz18} in a number of crucial ways. First, in their analysis $\mu_2$ is absent since they incorrectly assumed that $\langle\Gamma^3\rangle=0$. The fact that $\mu_2\neq 0$ means that the leading order HI contribution associated with particle inertia occurs at $O(St)$, not at $O(St^2)$ as claimed in \cite{yavuz18}. Moreover, since $\langle\Gamma^3\rangle<0$, then $\mu_2<0$, meaning that the leading order effect of particle inertia is to \emph{reduce} the clustering associated with HI. Second, in the study of \cite{yavuz18} they do not directly compute or estimate $\mu_4$, but rather infer its behavior indirectly by fitting their RDF expression to their experimental data. With this procedure they find that $\mu_3>0$. However, the theory shows that $\mu_3<0$, if we assume $\langle \Gamma^4_1\rangle$, $\langle \Gamma^4_2\rangle$, $\langle \Gamma^4_3\rangle$ are all positive, as seems reasonable (see also below). As discussed in the paper, we suggest that this is because the behavior they observed in their experiment is not described by the theory, and hence the coefficient inferred from their data does not in fact correspond to $\mu_3$.

Finally, we note that we could have split up the correlations involving $\Gamma^p_{ij}$ into separation contributions from the strain-rate and vorticity, as was done in \cite{chun05}. This allows the theory to capture the influence of the different Lagrangian timescales associated with the strain-rate and vorticity fields \cite{ireland16a}. However, this would have further complicated out result and would not change the conclusions drawn from our analysis.

\subsection{Computing the exponents using DNS data}

In order to quantitatively assess the predictions from the theory, we use our DNS data \cite{ireland16a} to compute the fluid statistics that appear in the definitions of $\mu_1, \mu_2, \mu_3, \mu_4$, and also use $B^{nl}=0.056$ as given in \cite{bragg14b}. With these we find $\mu_1\approx 31.98$, $\mu_2 \approx -0.02$, $\mu_3\approx -40.15$, $\mu_4\approx 15.76$.

The value $\mu_2 \approx -0.02$ is very small, and as a consequence the $\mathcal{O}(St)$ contribution in \eqref{RDF_sol} is only the leading order inertial correction to the HI induced clustering if $St<\mu_2/\mu_3\approx 6.14\times 10^{-4}$, i.e. extremely small. Yavuz et al claimed that $\langle\Gamma^3\rangle=0$ for an isotropic flow, and therefore $\mu_2=0$. We argued that this is incorrect, and our DNS supports this showing $\langle\Gamma^3\rangle=-0.15\tau_\eta^{-3}$. Instead, the reason why $\mu_2$ is small is because according to our DNS, $\tau_\Sigma=0.0034\tau_\eta$. This occurs because the associated autocorrelation function passes through zero at time lag $\approx 2.4\tau_\eta$ and then exhibits a significant negative loop, leading to a small value for the correlation timescale $\tau_\Sigma$.

The DNS data also confirms our expectation that $\mu_3$ is negative, demonstrating that the positive exponent measured in \cite{yavuz18} by fitting the RDF expression to their experimental data cannot be associated with $\mu_3$.

\begin{figure}[h]
\centering
\subfloat[]
{\begin{overpic}
[trim = 10mm 60mm 0mm 20mm,scale=0.3,clip,tics=20]{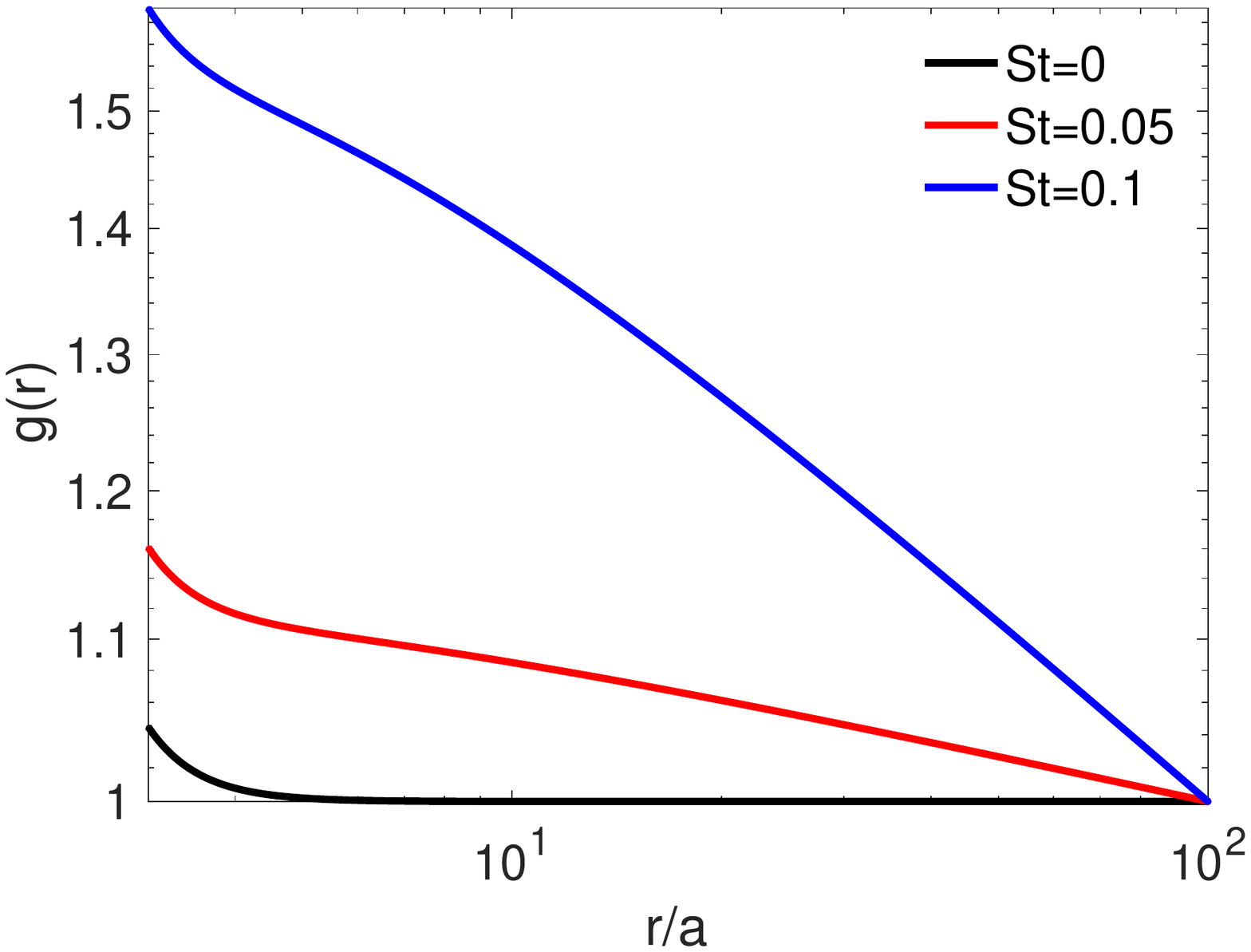}
\end{overpic}}
\vspace{-4mm}\subfloat[]
{\begin{overpic}
[trim =  10mm 60mm 0mm 20mm,scale=0.3,clip,tics=20]{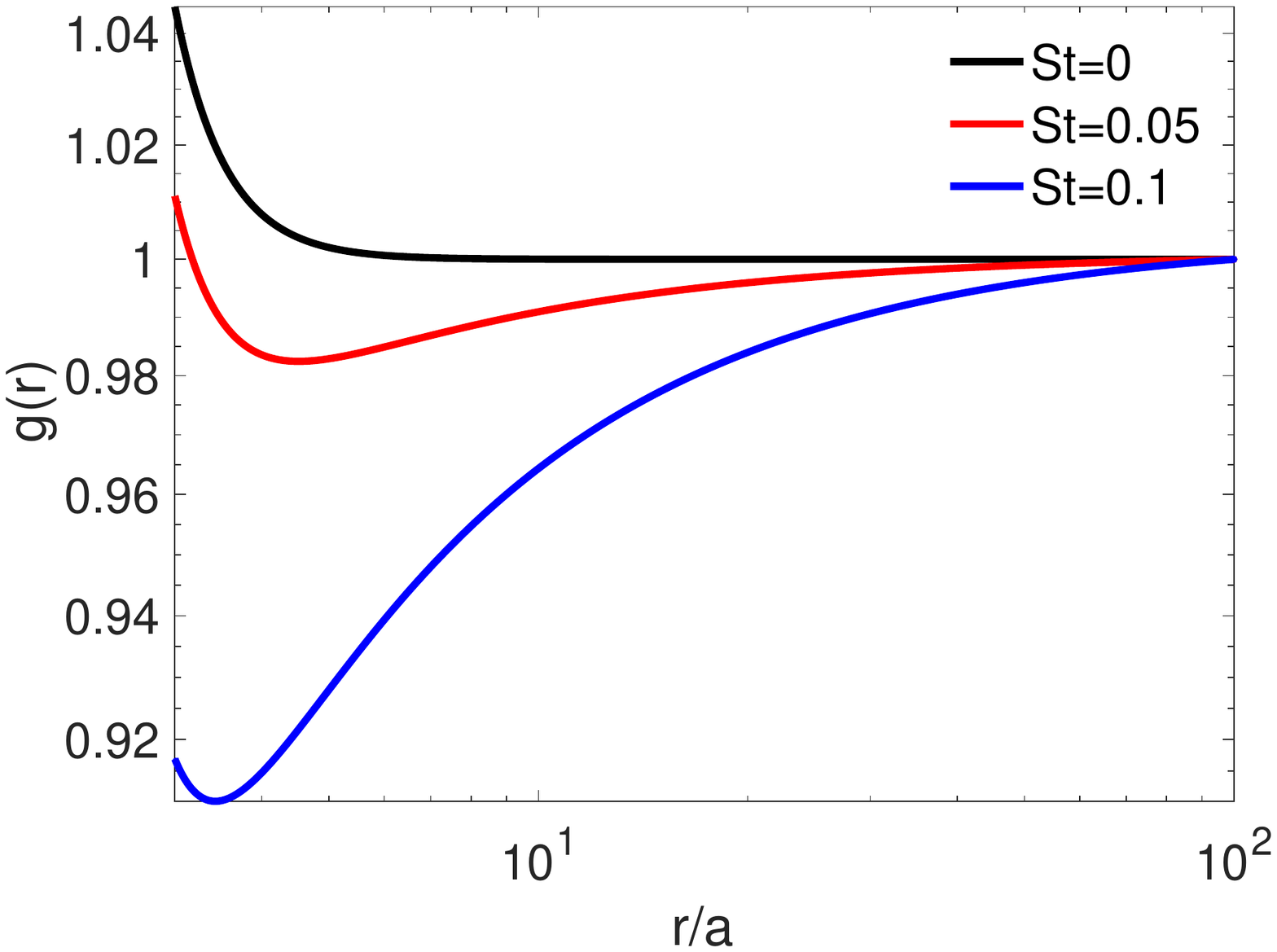}
\end{overpic}}
\caption{Plot of the theoretical predictions for the RDF, using the DNS data to prescribe $\mu_1, \mu_2, \mu_3, \mu_4$. Plot (a) is the full prediction including all of the terms in \eqref{RDF_sol}. Plot (b) uses $\mu_4=0$ in order to consider the role of the contributions that arise solely due to HI.} 
\label{RDF_plot}
\end{figure}

The predictions from the theory (assuming boundary condition $g(r=100a)=1$) are shown in figure~\ref{RDF_plot} for the far-field regime $r\gtrsim 3a$ for which they are supposed to be valid. Figure~\ref{RDF_plot}  (a) clearly shows that while inertia enhances the RDF, it does not enhance it by orders of magnitude. Furthermore, figure~\ref{RDF_plot} (b) shows the predictions when using $\mu_4=0$ in order to consider the role of the contributions that arise solely due to HI. From this it is clear that the inertial modification of the HI effects act to reduce the clustering compared to the $St=0$ case, rather than enhance it. Therefore, the enhancements due to inertia observed in figure~\ref{RDF_plot} (a) are solely due to the term involving $\mu_4$ which describes the clustering that arises even in the absence of HI, and is associated with the centrifuge effect and the associated preferential sampling of strain over rotation along the particle trajectory \cite{chun05}.

\section{Experiments}

\subsection{Error bars and uncertainty}
The uncertainties have been discussed in detail in \cite{hammond21}. For the uncertainty in $r$, the primary source of error arises from the use of track interpolation to identify particle position and thus particle-pair separation. This uncertainty appears since the non-zero radial relative velocity may change the instantaneous value of $r$ in the time between laser pulses. If fluctuations in $r$ occur that are over shorter timescales than the time between frames $\Delta t_2$, these fluctuations will not be recorded in the track. As such, we take the product of the standard deviation of radial relative velocity and $\Delta t_2$ to estimate the range of separations which may contribute to the recorded data at the given datapoint.

For the uncertainty in RDF, we found that the random error (quantified by the standard error) was extremely small ($<2\%$), since the data was well-converged. The potential for variation in RDF instead arose instead in the selection of inputs for STB. As such, we varied the most important, consequential input parameter in STB, the maximum allowable triangulation error $\epsilon$ by $\pm 10\%$, and took twice the standard deviation of the resulting RDFs as the vertical error bar in RDF.

In Fig. \ref{RDFErrBar}, we plot the interpolation uncertainty (in $r$) as a shaded region for the middle-most $St$ ($St=0.16$ in Fig. \ref{RDFErrBar}a and $St=0.74$ in Fig \ref{RDFErrBar}b) in both plots of the experimental results. Similarly, we plot the RDF uncertainty by ensemble forecast as vertical error bars for the middle-most $St$. As a result, we find that the variations among the different $St$ results across $r$ are within the uncertainty limits, but there are clear, weak positive trends in RDF with $St$. Additionally, the comparison of the uncertainty in $r$ of the $St=0.74$ case in Fig. \ref{RDFErrBar}(b) with the results of the main letter shows that for across the four particle types, the results of the $a= 8.75\mu m$, $14.25\mu m$, and $20.75\mu m$ particles generally collapse over decreasing $r$ in the explosive $r^{-6}$ regime.

{\vspace{0mm}\begin{figure}[h]
		\centering
		\subfloat[]
		{\begin{overpic}
				[scale=0.7,clip,tics=20]{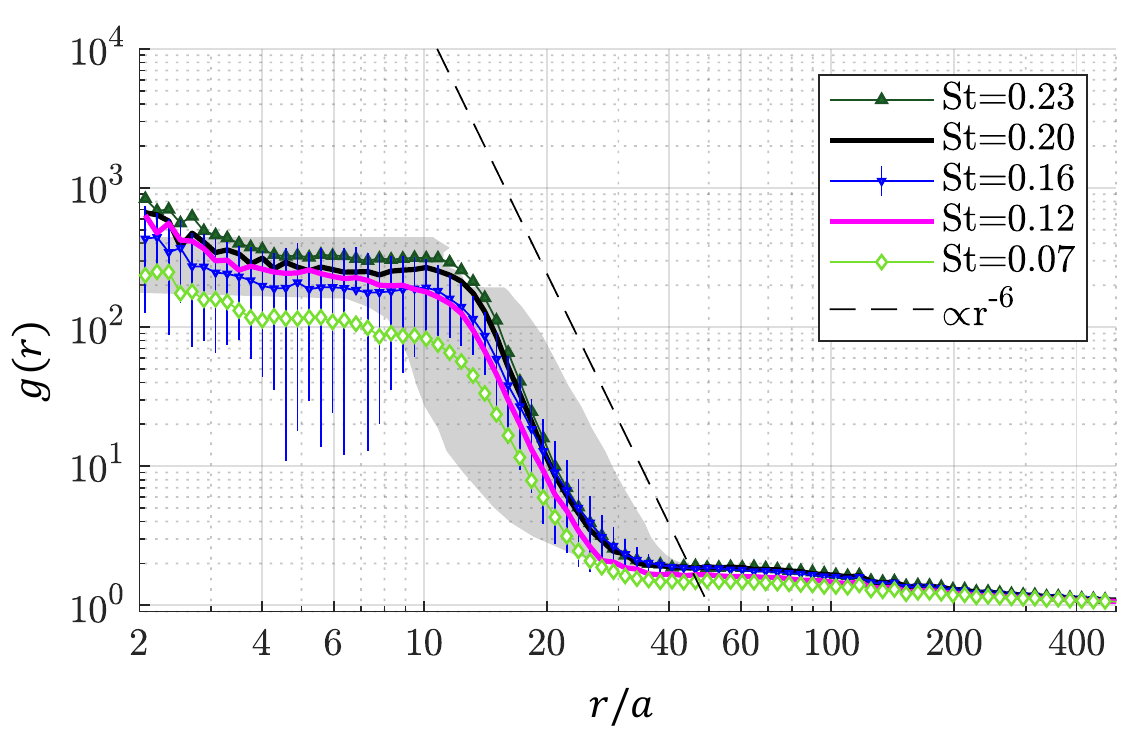}

		\end{overpic}}
		\hspace{4mm}\vspace{-4mm}\subfloat[]
		{\begin{overpic}
				[scale=0.7,clip,tics=20]{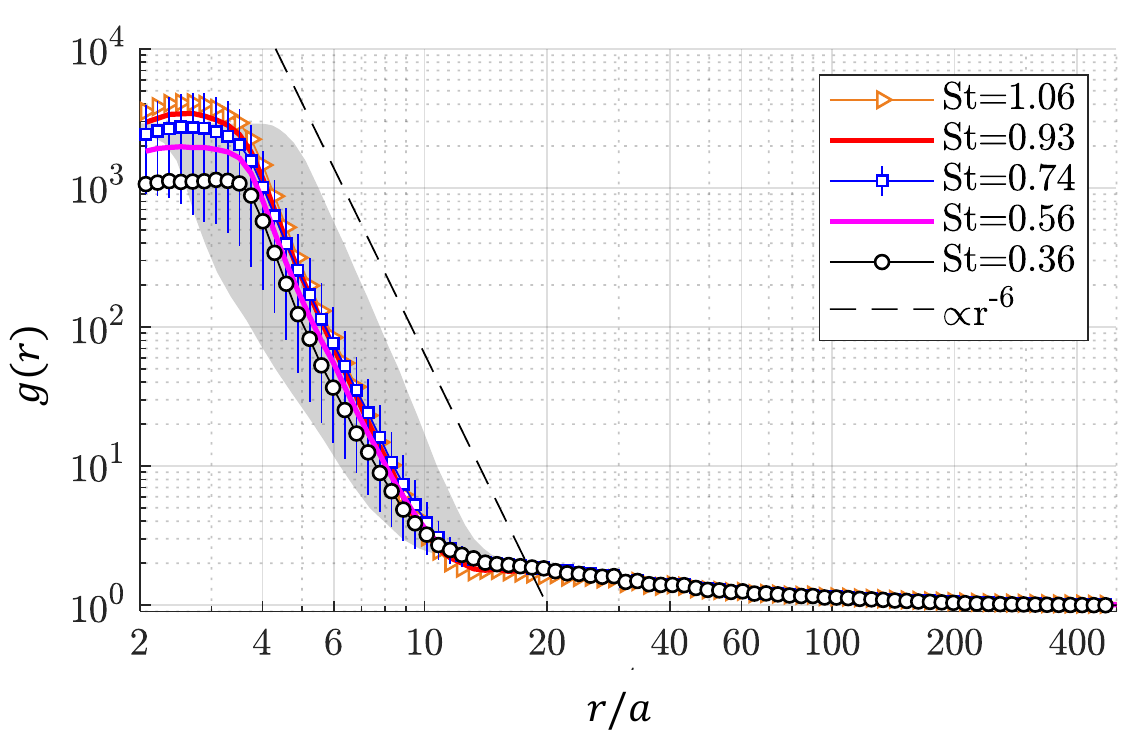}

		\end{overpic}}\\
    	\vspace{2mm}
		\caption{RDF for different $St$ and radius (a) $a=3.75\mu m$,  (b) $a=14.25\mu m$. The shaded regions represent horizontal error bars by interpolation uncertainty for the (a) $St=0.16$ case, and (b) $St=0.74$ case. The blue vertical error bars represent the RDF uncertainty for the (a) $St=0.16$ case, and (b) $St=0.74$ case.} 
		\label{RDFErrBar}
\end{figure}}

\subsection{Particle photographs to test for agglomeration}

To test if some of the particles inside the flow facility had agglomerated, we sampled particles from the HIT chamber, and took images using a microscope. The particles were sampled by applying an adhesive to a glass microscope slide, and inserting it into the flow facility during fan operation while seeded with the $a=21\mu m$ particles. A crop of this microscope image is shown in Fig. \ref{PtlIm}.

{\vspace{0mm}\begin{figure}
		\centering		
		\includegraphics[trim = 0mm 0mm 0mm 10mm,scale=1.7,clip,tics=20]{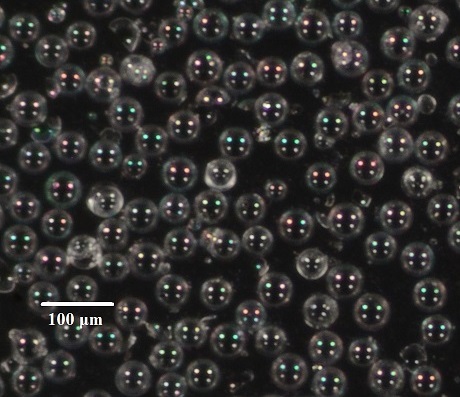}		
		\caption{$20.75\mu m$ particles sampled from the HIT chamber during operation.} 	
		\label{PtlIm}
\end{figure}}

Of 1684 uniformly sampled particles from the photograph, 87\% were regular spheres, while 7\% were double spheres (two spheres connected in the manner two soap bubbles touch, with varying individual radii) and 6\% were cracked or broken. As stated in the Letter, due to the enormous difference between the theoretical prediction and the experimental results, it is not expected that the 13\% of the particles that deviated from a regular spherical shape solely led to the observed difference between theory and experiments.

\subsection{Electric charge effect on the RDF experiments}

It was argued in the main text that electric charge does not play an important role in the particle clustering we observe. Here, we support this argument with experimental measurements of charge in our charge-minimized flow facility, where we find the Coulomb force is negligible even at the smallest observable separation $r\approx 2.07a$, and thus cannot be responsible for the observed extreme clustering. 

Particle charging can potentially occur in our facility by friction charging due to particle contact with the fans and inside walls of the flow chamber, a phenomenon known as the triboelectric effect. The magnitude of the charge generated by this process depends on the difference in the work functions of the two materials \cite{matsusaka10}. To prevent particle charging, we coated the flow facility fans and walls in carbon conductive shielding paint (Stewart Macdonald), with work function $\sim 5$ eV \cite{fomenko2012handbook,michaelson1977work}, to match the work function of silicon dioxide, the primary compound of the particles \cite{fomenko2012handbook}. The flow facility was grounded, such that the conductive coating would mitigate residual or pre-existing particle charge. 

To test if this material combination produced minimal charge, we measured particle charge in a small-scale turbulent flow facility \cite{tripathi2015improved}. The experiment was performed on the $a=14.25 \mu m$ glass bubble particles inside of a turbulent impinging flow tube with identical fans and surface coatings to the HIT chamber. We found that the resulting charge distribution was bipolar, with a mean of $3.5\times 10^{-17} \textrm{C}$ (C: Coulomb) and standard deviation $4.0\times 10^{-16} \textrm{C}$. 

To verify that the level of charge was similar in our full-scale HIT chamber, we also measured the electric charge on particles in situ. We used our charge measurement technique described in \cite{hammond2019holographic}, modified to observe deflection of particle trajectories in a strong electric field using Shake-the-Box particle tracking. By comparing the particle terminal velocity before and after the introduction of an electric field, we estimated the electric drift velocity $v_x$ for charge measurements \cite{hammond2019holographic}. We found that even when initially detectably charged particles were injected into the flow facility ($\approx 1\times 10^{-14} \textrm{C}$), after a time duration shorter than the startup time of the experiments ($\sim 30$s), particle charge was undetectable. We have calculated the resolution limit of this in-situ charge measurement system as $1.2\times 10^{-15} \textrm{C}$, based on the particle position resolution of 0.5 pixels. The electric charge is expected to be below this resolution limit.

To test if particle charge at the level of $10^{-15}$C could explain the extreme clustering observed, we compared the measured relative inward velocities of the particles with that which would be expected for oppositely-sign charged particles under idealized Coulomb attraction with charge $q=10^{-15}$C. In particular, we consider Coulomb attraction between two low-Reynolds-number particles (subject to Stokes drag) with charge magnitude $q$ and opposite charge sign, for which the magnitude of their inward relative velocity would be
\begin{equation}
w(r) = \frac{k_e q_1 q_2}{6\pi \mu a} \frac{1}{r^2} \: ,\label{eqchg}
\end{equation}
where $k_e$ is Coulomb's constant and $\mu$ is the dynamic viscosity of the fluid. In \cite{hammond21}, we have reported inward relative velocity statistics as a function of $r$ for the $St=0.74$, $a=28.5\mu m$ particles used in this study. If $q\approx10^{-15}$C, then \eqref{eqchg} predicts relative velocities that are smaller by two orders of magnitude compared to the measurements. In the regime where $g(r)-1\propto (r/a)^{-6}$, this gap widens to three orders of magnitude. Since the predicted relative velocity due to the Coulomb force between two particles with the largest possible electric charge is 3 orders of magnitude smaller than the average observed relative velocities, this offers additional evidence that the Coulomb force is weak at the observed spatial scales, and not an explanation for the extreme clustering observed in the experiments.

\bibliographystyle{unsrt}
\bibliography{SM}